\documentclass[lettersize,journal]{IEEEtran}
\usepackage{amsmath,amsfonts}
\usepackage{array}
\usepackage[caption=false,font=normalsize,labelfont=sf,textfont=sf]{subfig}
\usepackage{textcomp}
\usepackage{stfloats}
\usepackage{url}
\usepackage{verbatim}
\usepackage{graphicx}
\usepackage{cite}
\usepackage{multirow}
\usepackage{cite}  
\usepackage{bm}
\usepackage{amsmath}
\usepackage{booktabs}
\usepackage[linesnumbered,ruled]{algorithm2e}
\usepackage{algpseudocode}

\usepackage[colorlinks=true,
            linkcolor=blue,  % \ref, \eqref 的颜色 (公式、章节、图表)
            citecolor=blue,  % \cite 的颜色 (参考文献)
            urlcolor=blue    % \url 的颜色 (网页链接)
            ]{hyperref}

\makeatletter
\newcommand{\removelatexerror}{\let\@latex@error\@gobble}
\makeatother

\SetKwRepeat{Do}{do}{while}

\begin{document}
\begin{sloppypar}

\title{PGUDA: Pressure-Guided Unsupervised Domain Adaptation with Cross-Modal Knowledge Distillation for sEMG-Based Gesture Recognition}

\author{Yurui Liu\textsuperscript{\#}, Xiao-Cong Zhong\textsuperscript{\#}, Qisong Wang, Xuefu Wang, Dan Liu, and Jinwei Sun

\thanks{\textsuperscript{\#}These authors contributed equally to this work.}
\thanks{This work was sponsored by the National Natural Science Foundation of China (Grant No. 61471140, 62303420), and the Fundamental Research Funds for the Central Universities (Grant No. IR2021222), Future Science and Technology Innovation Team project of HIT (216506). (\textit{Corresponding authors: Qisong Wang.})}
\thanks{Yurui Liu, Xiao-Cong Zhong, Dan Liu, Jinwei Sun, and Qisong Wang are with the School of Instrumentation Science and Engineering, Harbin Institute of Technology, Harbin 150001, China (e-mail: liuyurui@stu.hit.edu.cn; zhongxiaocong@hit.edu.cn; wangxuefu@stu.hit.edu.cn; liudan@hit.edu.cn; jwsun@hit.edu.cn; wangqisong@hit.edu.cn).}
\thanks{ }
}

\markboth{Journal of \LaTeX\ Class Files,~Vol.~14, No.~8, August~2021}%
{Shell \MakeLowercase{\textit{et al.}}: A Sample Article Using IEEEtran.cls for IEEE Journals}

\maketitle

\begin{abstract}
Surface electromyography (sEMG)-based gesture recognition has emerged as a promising technology for natural human-computer interaction. 
However, its practical deployment remains challenging due to severe performance degradation caused by feature distribution discrepancies across different subjects and recording sessions. 
Although domain adaptation (DA) techniques are commonly employed to mitigate such discrepancies, conventional methods often struggle to effectively aligning sEMG features, primarily due to their inherent stochasticity and the scarcity of labeled data. 
To address these limitations, this paper proposes a novel Pressure-Guided Unsupervised Domain Adaptation (PGUDA) framework, which leverages the robustness and stability of pressure signals to introduce a cross-modal knowledge distillation strategy that transfers consistent physical semantics across modalities. Specifically, a teacher network trained on pressure signals guides an sEMG student network on unlabeled target domains, thereby regularizing the representation learning process with transferable and modality-invariant knowledge.
Extensive experiments conducted on a self-collected multimodal dataset involving eleven subjects validate the effectiveness of the proposed PGUDA framework. The results demonstrate that our proposed PGUDA achieves leading performance in both cross-subject and cross-session classification tasks, achieving average accuracies of 58.08\% and  substantially outperforming existing DA approaches. 
Notably, PGUDA exhibits remarkable label efficiency: it attains classification accuracy comparable to fully supervised benchmarks while requiring only 5\% of labeled data for teacher network training. This framework offers a robust and data-efficient solution that can significantly reduce the calibration burden in practical sEMG-based gesture recognition systems.
\end{abstract}


\section{Introduction}
\IEEEPARstart{H}{and} gesture recognition technology serves as a critical technology in human-computer interaction, prosthetic control, and rehabilitation engineering \cite{Gu2022}, \cite{Ni2024}. 
Among various physiological modalities, surface electromyography (sEMG) has gained widespread adoption due to its non-invasive nature, low latency, and capacity to capture muscle activation patterns associated with movement intent. Recent progress in deep learning has further enabled effective decoding of sEMG signals for complex gesture recognition, as demonstrated by several advanced models \cite{Wei2019}, \cite{Guo2023}, \cite{lai2021stcn}.

\begin{figure}
    \centering
    \includegraphics[width=1\linewidth]{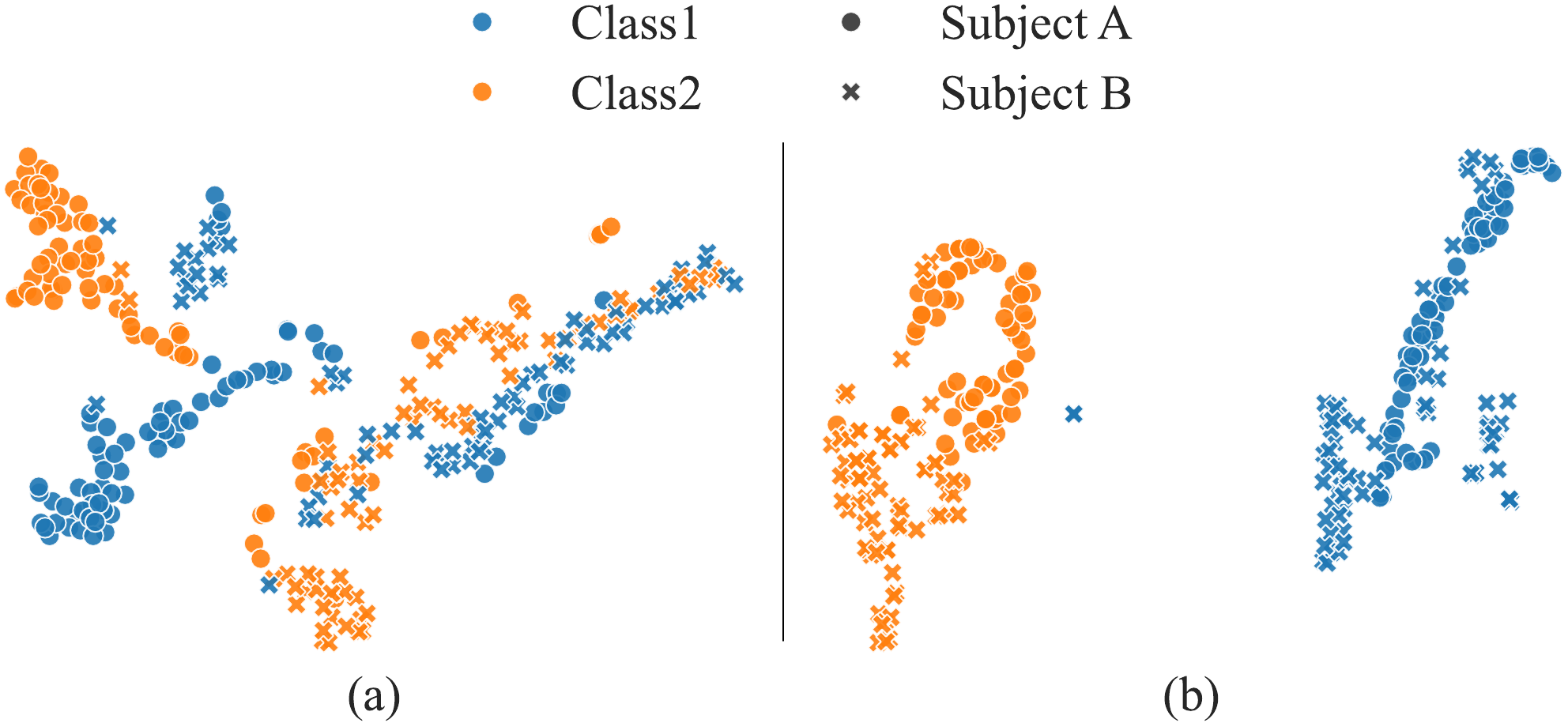}
    
    \caption{t-SNE visualization of feature distributions for different physiological modalities and subjects. (a) sEMG features; (b) Pressure features.}
    \label{fig:tsne1}
\end{figure}

However, the widespread deployment of sEMG-based systems is hindered by significant generalization challenges, especially in cross-subject and cross-session scenarios.
Classification models trained on source data often suffer from severe performance degradation when applied to a new target domain. 
This decline is mainly caused by inherent domain discrepancy, as illustrated in Fig. \ref{fig:tsne1}\hyperref[fig:tsne1]{(a)}, which stems from individual anatomical differences, such as muscle geometry, subcutaneous fat thickness, temporal non-stationarity due to factors like electrode displacement, skin impedance variation, and muscle fatigue \cite{Xiong9346072}.
Consequently, maintaining high recognition accuracy typically requires collecting substantial labeled data from each target user, a process that is labor-intensive and heavily dependent on dense annotations, thereby impeding practical usability and user acceptance.
% \IEEEpubidadjcol

To mitigate the reliance on target-domain labels, transfer learning techniques, particularly domain adaptation (DA) \cite{Kouw2021}, have been developed to align feature distributions across different domains \cite{DANN2016, Li2024}. Despite DA methods have achieved notable success in fields such as computer vision, their application to sEMG signals remains challenging. Unlike structured visual images, sEMG signals are inherently stochastic, susceptible to environmental noise, and lack visually interpretable patterns. 
As a result, directly minimizing domain discrepancy without sufficient semantic guidance often leads to negative transfer, where samples from distinct gesture classes become erroneously aligned due to poorly defined decision boundaries in the target domain.
Meanwhile, while cross-modal knowledge distillation (CMKD) has demonstrated effectiveness in integrating heterogeneous signals \cite{7780678}, \cite{Gou2020KnowledgeDA}, its potential for addressing domain shift and label scarcity in unsupervised sEMG-based adaptation remains largely unexplored.
 
To address these challenges, this paper proposes a Pressure-Guided Unsupervised Domain Adaptation (PGUDA) framework with cross-modal knowledge distillation for sEMG-based gesture recognition. 
The design is motivated by the observation that tactile pressure signals captured from fingertips directly reflect the mechanical outcomes of hand gestures. As illustrated in Fig. \ref{fig:tsne1}, compared with sEMG modality, pressure data exhibit notably greater intra-class compactness, wider inter-class separation, and substantially reduced domain discrepancy. Consequently, pressure signals demonstrate higher inherent stability and robustness against bio-electrical noise. 
Although pressure sensors are generally less practical for long-term daily use, they can serve as an ideal teacher modality during offline training. 
By leveraging cross-modal knowledge distillation, the proposed method transfers robust, domain-invariant knowledge from the pressure modality to the sEMG student network. This semantic guidance acts as a reliable anchor that regularizes the representation learning process and enhances recognition accuracy in unlabeled target domains.
The main contribution of this article are summarized as follows:
\begin{itemize}
    \item We propose a novel multimodal unsupervised domain adaptation framework, called PGUDA, which utilizes pressure signals to guide sEMG feature learning, effectively addressing both domain discrepancy and label scarcity in sEMG-based gesture recognition.
    \item We introduce a cross-modal knowledge distillation mechanism that leverages pressure signals as robust semantic guidance to enhance the noise and non-stationarity robustness of sEMG representations, while an alignment module concurrently minimizes global inter-domain distribution discrepancy.
    \item Extensive experiments are conducted under cross-subject and cross-session protocols, demonstrating that our proposed PGUDA achieves greater performance and exhibits superior label efficiency compared to existing domain adaptation and supervised methods. 
    % attaining performance comparable to supervised baselines with only 5\% of labeled data.
\end{itemize}

The remainder of this paper is organized as follows. Section II reviews the related works. Section III details the methodology of the proposed PGUDA framework. Section IV describes the experimental setup and presents a comprehensive analysis of the results. Section V provides a discussion, and Section VI concludes the paper and identify future directions.

\section{Related Works}
\subsection{sEMG-based Gesture Recognition}
Early research on sEMG-based gesture recognition relied primarily on handcrafted feature extraction coupled with conventional machine learning algorithms. Englehart \textit{et al.} \cite{Englehart2003} introduced a robust method utilizing linear discriminant analysis (LDA), demonstrating its effectiveness in processing four-channels sEMG signals. Oskoei \textit{et al.} \cite{Oskoei2008} conducted a comprehensive evaluation showing that support vector machines (SVM) outperform linear classifiers in handling the high-dimensional and nonlinear characteristics of sEMG patterns. More Recently, the rise of deep learning has fundamentally transformed this field by enabling end-to-end automatic feature learning. Guo \textit{et al.} \cite{Guo2023} employed 1D-CNN and InceptionTime architecture for classifying ten hand gestures, achieving test accuracies of 80.98\% and 90.89\%, respectively. Lai \textit{et al.} \cite{lai2021stcn} designed a spatial-temporal convolutional network (STCN) to capture both spatial and temporal correlations within sEMG signals, reporting accuracies of 99.8\% on the CapgMyo dataset and 75.8\% on the BandMyo dataset. Burrello \textit{et al.} \cite{burrello2022bioformers} proposed Bioformers, a lightweight transformer-based model, which attained 65.73\% accuracy on the Ninapro DB6 dataset with eight gestures while drastically reducing computational cost for ultra-low-power inference. 

Despite these advances, most existing methods remain limited in generalizing across different subjects and sessions, and are highly dependent on abundant labeled data from new users. These limitations severely hinder their robustness and practical deployment in real-world applications.

\subsection{Cross-Modal Knowledge Distillation}
Knowledge distillation, originally proposed by Hinton \textit{et al.} \cite{hinton2015distilling}, is a learning paradigm in which a compact student model learns to mimic the soft output distributions of a larger teacher model. Initially developed for model compression, this concept has been extended to cross-modal knowledge distillation (CMKD), where knowledge is transferred across different sensory modalities \cite{Gou2020KnowledgeDA}. For example, Gupta \textit{et al.} \cite{7780678} demonstrated that supervision from a labeled teacher modality can effectively guide a student model trained on a distinct unlabeled modality, substantially enhancing the student's performance.

In the domain of physiological signal processing, CMKD has shown notable success in integrating heterogeneous data streams. For emotion recognition, Liu \textit{et al.} \cite{EmotionKD} proposed EmotionKD and Kan \textit{et al.} \cite{kan2025CMCRD} introduced cross-modal contrastive representation distillation, both of which leverage multimodal teachers to guide unimodal students and reduce the inherent noise and instability of physiological signals. Similarly, in continuous sign language recognition, Gao \textit{et al.} \cite{GAO2024106587} employed CMKD to transfer semantic knowledge from visual and textual modalities to sign gloss sequences, improving robustness against complex hand movements. In sEMG-based gesture recognition, Zeng \textit{et al.} \cite{Zeng9845471} utilized A-mode ultrasound (AUS) to supervise an sEMG network, demonstrating that anatomical knowledge distilled from AUS can significantly enhance sEMG decoding accuracy. Additionally, Li \textit{et al.} \cite{Li2024SelfDistillation} explored multi-task learning combined with self-distillation to improve gesture recognition efficiency on edge devices.

Despite these advances, existing CMKD studies in sEMG recognition have primarily focused on within-domain settings, largely overlooking the critical issue of domain discrepancy across subjects and sessions. To bridge this gap, we propose a pressure-guided strategy that exploits the physical stability of tactile pressure signals to supervise sEMG feature adaptation. This method effectively restores recognition performance in unlabeled target domains, eliminating the need for explicit user calibration or annotation.

\subsection{Domain Adaptation}
Domain adaptation has been widely studied to overcome the domain discrepancy in sEMG-based gesture recognition. For instance, Wang \textit{et al.}\cite{WANG2024106086} and Li \textit{et al.} \cite{LI2025106892} utilized DA techniques such as correlation alignment (CORAL) \cite{Zhong2023} to reduce feature distribution discrepancies between source and target domains, improving cross-domain generalization. Zhong \textit{et al.} \cite{Zhong2025} proposed a plug-and-play subdomain adaptation method (PPSDA) that uses local maximum mean discrepancy to refine alignment within discriminative subdomains, addressing structural confusion common in global adaptation strategies. Côté-Allard \textit{et al.} \cite{Côté2021Gesture} proposed transferable adaptive domain adversarial neural networks, which learn subject-invariant representations by training a feature extractor to confuse a domain discriminator. Su \textit{et al.} \cite{Su10971950} developed a multisource adversarial feature disentanglement method to explicitly separate domain-invariant features from subject-specific noise. Additionally, Fratti \textit{et al.} \cite{Frattis24227147} demonstrated the effectiveness of transfer learning via fine‑tuned multi‑scale CNNs, adapting pre‑trained models to specific prosthetic control tasks.

Beyond unimodal adaptation, recent research in other fields has shifted toward leveraging complementary modalities to bridge the domain gap. Bhalla \textit{et al.} \cite{Bhalla2022IMU2Doppler} introduced a cross-modal framework that transfers knowledge from inertial measurement units (IMUs) to Doppler radar, enabling label‑free activity recognition in the target domain. In 3D semantic segmentation, Jaritz \textit{et al.} \cite{Jaritz2023Semantic} proposed a consistency‑enforcing strategy between 2D images and 3D point clouds, where the more reliable modality guides adaptation on unlabeled target data.

Drawing on the insights from the aforementioned studies, we hold that sEMG-based gesture recognition can similarly benefit from guidance via a stable auxiliary modality. In this work, we leverage tactile pressure signals as a physically grounded anchor to guide the unsupervised adaptation of sEMG networks, ensuring semantic consistency across domains without reliance on target‑domain labels.

\begin{figure*}[t] % [t] 表示强制置顶，跨栏图通常只能放在顶部
    \centering

    \includegraphics[width=1\linewidth]{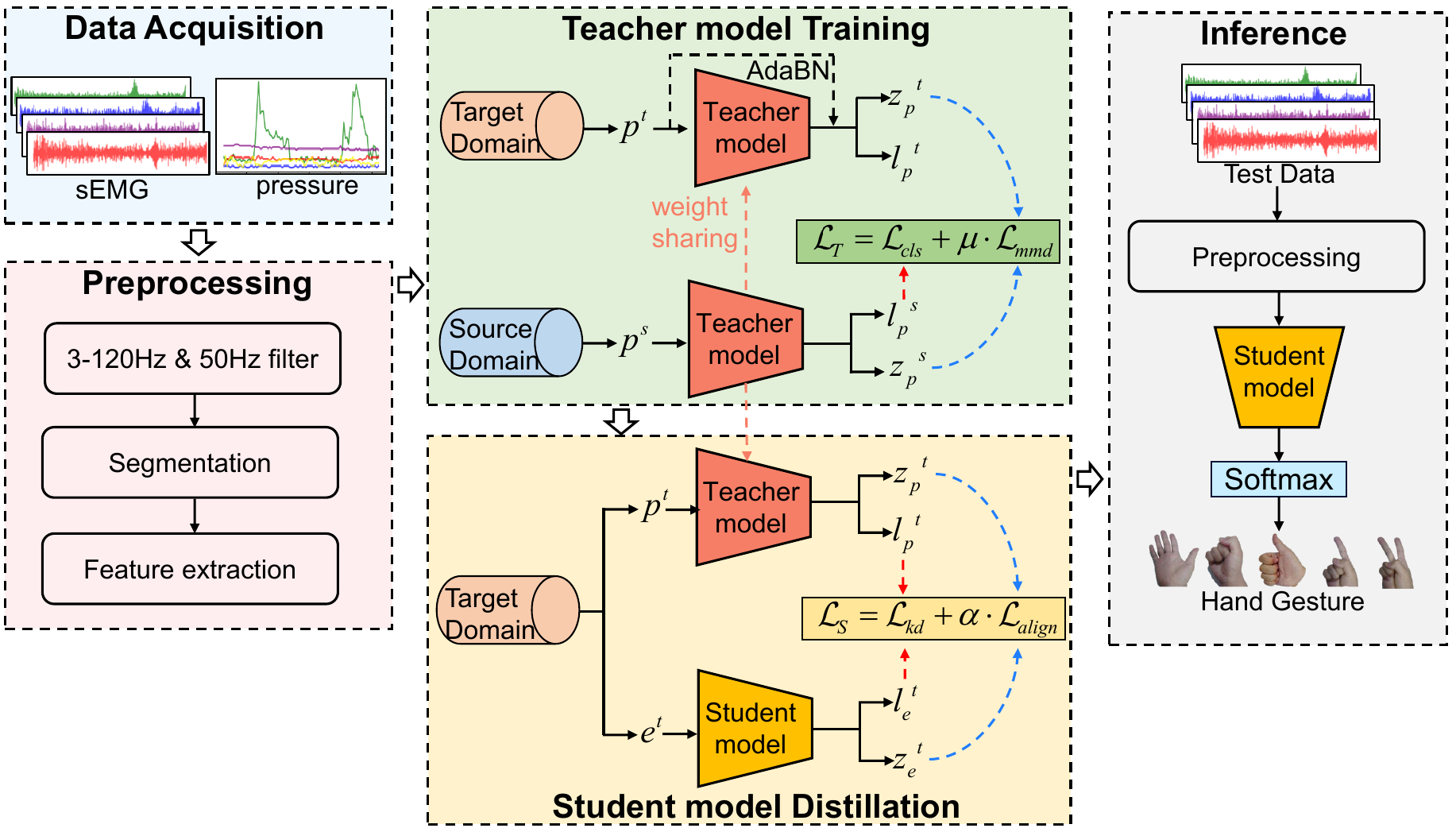}
    
    \caption{Overview of the proposed PGUDA framework. The left panel presents the multimodal data acquisition and preprocessing pipeline. The middle panel details the two-stage training strategy, including teacher model training and cross‑modal knowledge distillation to the student network. The right panel illustrates the inference stage, where only sEMG signals are used for gesture recognition.}
    
    \label{fig:framework}
\end{figure*}

\section{Methodology}
\subsection{Problem Formulation}
In this work, we address the problem of unsupervised domain adaptation (UDA) for sEMG-based gesture recognition under guidance from a pressure modality. Let $\mathcal{E}$ denote the sEMG signal space, $\mathcal{P}$ the pressure signal space, and $\mathcal{Y} = {1, 2, \dots, K}$ the label set of $K$ gesture classes. We define two distinct domains with different feature distributions as follows:
\begin{itemize}
    \item \textbf{\textit{Source domain} $\mathcal{D}_s$:} This domain comprises labeled data collected from a source subject or session. It consists of $N_s$ samples, denoted as $\mathcal{D}_s = \{(\bm{e}_i^s,\bm{p}_i^s,y_i^s)\}_{i=1}^{N_s}$, where $\bm{e}_i^s \in \mathcal{E}$ and $\bm{p}_i^s \in \mathcal{P}$ represent the sEMG and pressure samples respectively, and $y_i^s \in \mathcal{Y}$ is the corresponding ground‑truth gesture label.
    \item \textbf{\textit{Target domain} $\mathcal{D}_t$:} This domain consists of unlabeled data from a new target subject or session. It includes $N_t$ samples, expressed as $\mathcal{D}_t = \{(\bm{e}_j^t, \bm{p}_j^t)\}_{j=1}^{N_t}$, where gesture labels are unavailable during adaptation.
\end{itemize}

The core challenge stems from the probability distribution difference of sEMG signals between the source and target domains, formally expressed as $P(\bm{e}^s) \neq P(\bm{e}^t)$. To address this, we leverage the observation that pressure signals exhibit greater cross-subject and cross-session stability, which we formalize as $P(\bm{p}^s) \approx P(\bm{p}^t)$.

Our goal is to learn a robust sEMG student network $f_{\theta_S}(\cdot)$ that generalizes effectively the unlabeled target domain $\mathcal{D}_t$. Different from conventional UDA approaches, the student network is not directly supervised by source labels. Instead, the training process is structured in two decoupled stages:
\begin{itemize}
    \item  \textbf{\textit{Teacher Training}}: A teacher network $\bm{f}_{\theta_T}(\cdot)$ is first trained on the source domain using pressure signals and their corresponding ground-truth labels.
    \item  \textbf{\textit{Student Learning}}: The student network $\bm{f}_{\theta_S}(\cdot)$, which takes sEMG signals as input, is then guided by the teacher's soft predictions on the target domain, thereby regularizing its representation learning without access to target labels.
\end{itemize}
Mathematically, the optimization problem is formulated as minimizing a joint objective function that combines a supervised classification loss with a cross-modal domain adaptation loss:
\begin{equation}
    \label{eq:objective}
    \begin{split}
    \theta_S^* = \mathop{\arg\min}_{\theta_T,\theta_S} \bigg( & \frac{1}{N_s}\sum_{i=1}^{N_s} \mathcal{L}_T(\bm{f}_{\theta_T}(\bm{p}_i^s), y_i^s) \\
    & + \lambda \cdot \frac{1}{N_t}\sum_{j=1}^{N_t} \mathcal{L}_S(\bm{f}_{\theta_S}(\bm{e}_j^t), \bm{f}_{\theta_T}(\bm{p}_j^t)) \bigg),
    \end{split}
\end{equation}
where $\mathcal{L}_T(\cdot,\cdot)$ denotes the supervised classification loss for training the teacher network, and $\mathcal{L}_S(\cdot,\cdot)$ represents the cross-modal UDA loss that aligns the student predictions with the teacher's guidance on the target domain. During inference, only the optimized student model $\bm{f}_{\theta_S^*}(\cdot)$ is deployed, while the teacher network is no longer required.

\subsection{Data Segmentation and Feature Extraction}
To process continuous physiological signals, we employed a sliding window technique \cite{Xiong9346072}, segmenting both multi-channel sEMG and pressure signals into 200 ms windows with a stride of 100 ms. This configuration was chosen to achieve a trade‑off between information content and real‑time latency. Synchronization between the sEMG and pressure windows was strictly enforced via timestamp alignment.

Feature extraction plays a vital role in compressing high-dimensional raw data into discriminative representations for pattern recognition. Similar the practice in \cite{Wuapp9245343, Das10719910}, we constructed a hybrid sEMG feature set by combining time‑domain and frequency‑domain descriptors. Specifically, the extracted feature vector includes: root mean square (RMS), mean absolute value (MAV), variance (VAR), waveform length (WL), differential absolute standard deviation (DASDV), power spectral area (PSA), and maximum power spectrum (MPS). These features collectively characterize muscle activation patterns. Let $\bm{x}_i$ denote the $i$-th sample within a segmented window of length $N$. The mathematical definitions and physiological interpretations of the features are given below:
\begin{equation}
 \text{RMS} = \sqrt {\frac{1}{N}\sum\limits_{i = 1}^N {{\bm{x}_i}^2} },
\end{equation}
which reflects the signal’s steady‑state power and, in the context of sEMG, serves as an estimator of the amplitude of quasi‑constant force during non‑fatiguing contractions.
\begin{equation}
\text{MAV} = \frac{1}{N}\sum\limits_{i = 1}^N {|{\bm{x}_i}|} ,
\end{equation}
which calculates the average rectified value of the signal. This feature serves as a direct indicator of muscle contraction intensity and is widely used due to its computational efficiency.
\begin{equation}
\text{VAR} = \frac{1}{N}\sum\limits_{i = 1}^N {{{({\bm{x}_i} - \bm{\overline{x}} )}^2}} ,
\end{equation}
where $\bm{\overline{x}}$ denotes the mean of the segment. VAR measures the signal's deviation from its mean and is interpreted as the AC power of the sEMG, quantifying the degree of signal fluctuation.
\begin{equation}
\text{WL} = \sum\limits_{i = 1}^{N - 1} {|{\bm{x}_{i + 1}} - {\bm{x}_i}|},
\end{equation}
which computes the cumulative waveform length over the time window, implicitly capturing information related to signal complexity, amplitude, and frequency content.
\begin{equation}
\text{DASDV} = \sqrt {\frac{1}{{N - 1}}\sum\limits_{i = 1}^{N - 1} {{{({\bm{x}_{i + 1}} - {\bm{x}_i})}^2}} }, 
\end{equation}
which is a standard deviation measure applied to the difference between adjacent samples. It is particularly sensitive to high-frequency components of the sEMG signal.

To characterize spectral properties, the power spectral density $\text{PSD} (f_i)$ is obtained via the fast Fourier transform, where $f_i$ represents the $i$-th frequency bin.
\begin{equation}
\text{PSA} = \sum\limits_{i = 1}^N {{f_i} \cdot {\text{PSD}}({f_i})} ,
\end{equation}
which estimates the integral of the power spectrum, reflecting the overall energy distribution of the muscle activity across the frequency domain.
\begin{equation}
\text{MPS} = \max \big({\text{PSD}}({f_i})\big),
\end{equation}
which identifies the magnitude of the dominant frequency component, aiding in the discrimination of gestures with similar temporal patterns but distinct spectral signatures.

After extraction, these seven features are concatenated across all channels to form the sEMG feature vector $\bm{v}_e \in \mathbb{R}^{C_e \times 7}$, where $C_e$ denotes the total number of sEMG channels.
For the pressure modality, given its inherent low-frequency stability, we simply compute the mean amplitude of each window as a represent of contact force intensity. Formally, for the $k$-th pressure channel, the feature $\bm{v}_p^k$ is obtained as:
\begin{equation}
    \bm{v}_p^k = \frac{1}{N_p}\sum_{n=1}^{N_p}p^k[n],
\end{equation}
where $p^k[n]$ is the raw reading at the $n$-th time step. This yields a compact pressure feature vector $\bm{v}_p \in \mathbb{R}^{C_p \times 1}$, with $C_p$ being the total number of pressure channels. 

\subsection{Teacher Model Training}
As illustrated in Fig. \ref{fig:framework}, the proposed PGUDA framework employs a teacher-student architecture. The first training stage focuses on optimizing the teacher model using pressure data from both the source and target domains.

\subsubsection{Pressure Teacher Network}
Given the low dimensionality of pressure signals, the teacher network $f_{\theta_T}$ is implemented as a lightweight multi‑layer perceptron (MLP) that processes the pressure feature vector $\bm{v}_p \in \mathbb{R}^{C_p \times 1}$. The network comprises three fully‑connected layers with hidden dimensions $[64, 128, 64]$. Let $\bm{h}_l$ denote the feature representation at the $l$-th layer. The transformation to the next layer $\bm{h}_{l+1}$ is expressed as:
\begin{equation}
    \bm{h}_{l+1} =  \mathcal{F}  \left( \bm{W}_l \bm{h}_l + \bm{b}_l \right),
\end{equation}
where $\bm{W}_l$ and $\bm{b}_l$ represent the learnable weight matrix and bias, and $\mathcal{F}(\cdot)$ denotes a non‑linear activation unit composed of batch normalization, ReLU, and dropout.

The output of the last hidden layer, denoted as $\bm{z}_{p} \in \mathbb{R}^{64 \times1}$, serves as a latent embedding. This feature vector acts as a robust semantic anchor for guiding the sEMG student network during domain adaptation, while also being mapped to the label space via a linear classifier.

\begin{figure}
    \centering
    \includegraphics[width=1\linewidth]{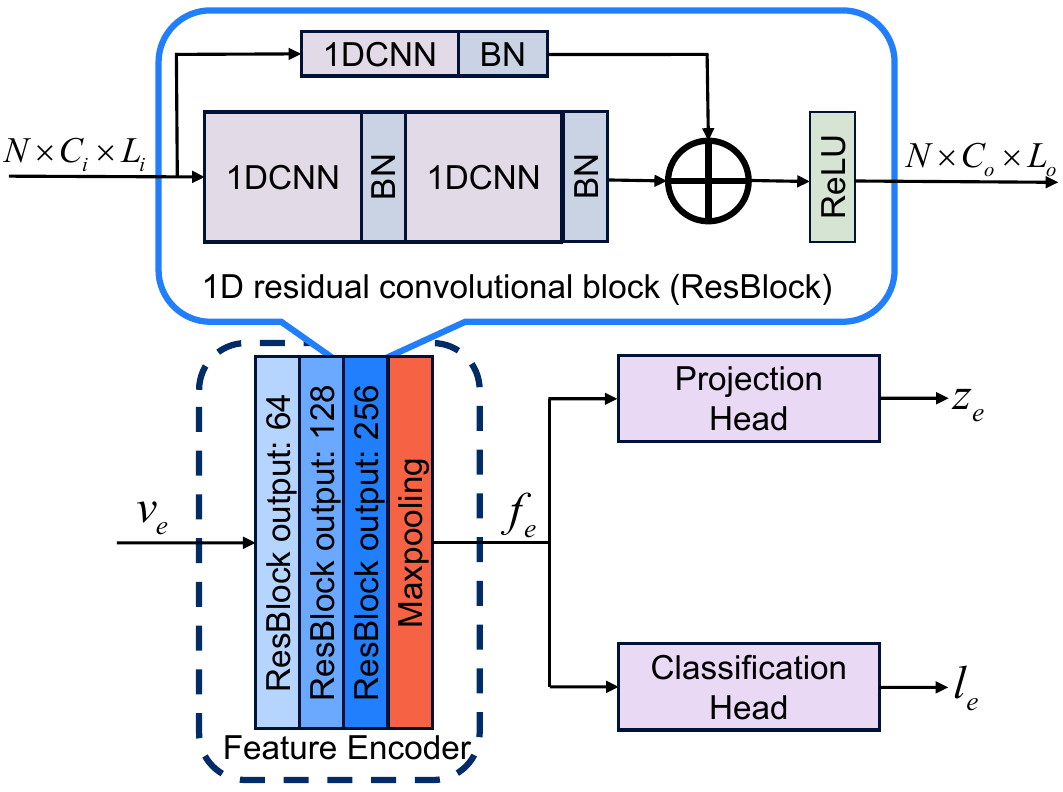}
    \caption{The architecture of the sEMG student network.}
    \label{fig:studentnetworks}
\end{figure}

\subsubsection{Training Strategy}
To obtain a reliable teacher model $\bm{f}_{\theta_T}(\cdot)$, we employ a hybrid adaptation strategy combining joint optimization and adaptive batch normalization (AdaBN)  \cite{Li2018}.
First, to ensure discriminative capability on the source domain, we minimize the cross-entropy loss $\mathcal{L}_{ce}$ on labeled pressure data. The supervised classification loss is formulated as:
\begin{equation}
\label{loss_cls}
\mathcal{L}_{cls} =  \frac{1}{N_s} \sum_{i=1}^{N_s} \mathcal{L}_{ce} (\bm{f}_{\theta_T}(\bm{p}_i^s), y_i^s).
\end{equation}
Simultaneously, to reduce the domain gap, we align the pressure feature distributions of source and target domains using maximum mean discrepancy (MMD) in a reproducing kernel Hilbert space (RKHS) $\mathcal{H}$:
\begin{equation}
\label{loss_mmd}
\mathcal{L}_{mmd} = \left\| \frac{1}{N_s} \sum_{i=1}^{N_s} \phi(\bm{z}_{p,i}^s) - \frac{1}{N_t} \sum_{j=1}^{N_t} \phi(\bm{z}_{p,j}^t) \right\|_{\mathcal{H}}^2,
\end{equation}
where $\phi(\cdot)$ denotes the kernel mapping function.
The overall objective for the first stage is therefore:
\begin{equation}
\label{eq:loss_T}
\min_{\theta_T} \, \mathcal{L}_{T} = \mathcal{L}_{cls} + \mu \cdot \mathcal{L}_{mmd},
\end{equation}
where $\mu$ is a hyper-parameter balancing the classification loss and the domain adaptation loss.

To further counteract against statistical discrepancies, we apply AdaBN as a post‑optimization refinement. Specifically, we freeze the learnable parameters of the teacher network and perform a forward pass over the target data, updating the weights  in all batch‑normalization layers. This explicitly adapts the teacher model to the target distribution without gradient‑based parameter updates.

\subsection{Student Model Distillation}
As shown in Fig. \ref{fig:studentnetworks}, the second stage performs knowledge distillation to transfer robust physical knowledge from the pressure teacher model to the sEMG student network.

\subsubsection{sEMG Student Network}
The student network $\bm{f}_{\theta_S}(\cdot)$ acts as the inference model. Architecturally, it is divided into a feature‑encoding backbone and two task‑specific heads. To capture complex temporal and channel-wise dependencies in the sEMG feature vectors $\bm{v}_e$, we adopt a 1D residual convolutional network as the backbone. Through stacked residual blocks, the backbone extracts a high‑level latent representation $\bm{f}_{e} \in \mathbb{R}^{256\times1}$ without suffering from gradient degradation.
After feature extraction, the network splits into two branches. The first is a projection head that maps $\bm{f}_e$ to a lower‑dimensional subspace $\bm{z}_e \in \mathbb{R}^{64\times1}$. Inspired by contrastive learning and domain‑adaptation methods \cite{CoteAllard2019, chen2020simclr}, we perform feature alignment in this subspace, which preserves intrinsic sEMG structure while mitigating domain discrepancy. The second branch is a classification head that directly maps $\bm{f}_{e}$ to gesture probabilities, focusing on discriminative performance. During inference, only the backbone and classification head are retained, ensuring low‑latency prediction.

\subsubsection{Distillation Strategy}
In this stage, the teacher parameters $\theta_T$ are frozen, and the student network $\bm{f}_{\theta_S}(\cdot)$ is optimized using unlabeled target data. We employ a dual‑level knowledge transfer that enforces consistency at both the probability and feature levels.

For class‑specific knowledge transfer, we apply cross-modal knowledge distillation via Kullback-Leibler divergence with a temperature scaling parameter $T$. Let $q_k(\cdot, T)$ denote the soft probability for class $k$:
\begin{equation}
q_k(\bm{l}, T) = \frac{\exp(\bm{l}_k / T)}{\sum_{j}^K \exp(\bm{l}_j / T)},
\end{equation}
where $\bm{l}$ is the output logits. The distillation loss $\mathcal{L}_{kd}$ is defined as:
\begin{equation}
\label{loss_kd}
        \mathcal{L}_{kd} = T^2 \sum_{k=1}^{K} q_k(\bm{l}_p^t, T) \log \left( \frac{q_k(\bm{l}_p^t, T)}{q_k(\bm{l}_e^t, T)} \right),
\end{equation}
where $\bm{l}_p^t$ and $\bm{l}_e^t$ are the logits generated by the teacher and student network on target data, respectively. The factor $T^2$ aligns the gradient magnitudes with loss scale.

Probability‑level alignment alone is insufficient for distribution matching. To enforce physical semantic consistency at the feature level, we minimize the MMD between the student's projected features $\bm{z}_e$ and the teacher's features $\bm{z}_p$. The alignment loss $\mathcal{L}_{align}$ is defined as: 

\begin{equation}
\label{loss_align}
    \mathcal{L}_{align} = \left\| \frac{1}{N_t} \sum_{j=1}^{N_t} \phi(\bm{z}_{e,j}^t) - \frac{1}{N_t} \sum_{j=1}^{N_t} \phi(\bm{z}_{p,j}^t) \right\|_{\mathcal{H}}^2.
\end{equation}
The overall objective for the student stage is:
\begin{equation}
\label{eq:loss_S}
\min_{\theta_S} \, \mathcal{L}_{S} =  \mathcal{L}_{kd} +\alpha \cdot \mathcal{L}_{align},
\end{equation}
where $\alpha$ controls the relative weight of feature‑level alignment.
The complete two‑stage training procedure of PGUDA is summarized in Algorithm \ref{alg:pguda}.
% 方案 A：如果你的算法比较窄，希望它待在双栏中的“单栏顶部” [t]（最推荐）
\begin{algorithm}[t]
    \caption{Training Procedure of the PGUDA Framework.}
    \label{alg:pguda}
    
    % \DontPrintSemicolon  % 如果你想去掉每行行末的分号，可以取消这行的注释
    \LinesNumbered        % 开启自动显示行号
    
    \SetKwInOut{Input}{Input}
    \SetKwInOut{Output}{Output}
    
    \Input{
        Source domain $\mathcal{D}_s$ and target domain $\mathcal{D}_t$; \\
        training epochs $E_1, E_2$; batch size $B$; \\
        hyper-parameters $\alpha, \mu$.
    }
    \Output{Classification results $\{y_j\}_{j=1}^{N_t}$ on $\mathcal{D}_t$.} 

    \BlankLine % 插入一个空白行，排版更美观
    Initialize parameters $\theta_T$ and $\theta_S$ \; 
    
    \textbf{\textit{Stage 1: Teacher model Training}} \; 
    
    \While{epoch $e < E_1$}{
        Obtain features and logits: \\
        $(\bm{z}_p^s, \bm{l}_p^s) \leftarrow f_{\theta_T}(\bm{p}^s)$, \\
        $(\bm{z}_p^t, \bm{l}_p^t) \leftarrow f_{\theta_T}(\bm{p}^t)$ \;
        
        Calculate $\mathcal{L}_{cls}$ and $\mathcal{L}_{mmd}$ by Eq.\eqref{loss_cls} and Eq.\eqref{loss_mmd} \;
        
        Update $\theta_T \leftarrow  \nabla(\mathcal{L}_{cls} + \mu \mathcal{L}_{mmd})$ by Eq.\eqref{eq:loss_T} \;
    }

    Freeze $\theta_T$, Set model to training mode \;
    
    \For{batch $\bm{p}_t$ in $\mathcal{D}_t$}{
        Forward pass $f_{\theta_T}(\bm{p}_t)$ to update BN statistics \;
    }
    Set calibrated teacher $\theta_T^* \leftarrow \theta_T$ \;

    \BlankLine
    \textbf{\textit{Stage 2: Student model Distillation}} \; 

    \While{epoch $e < E_2$}{
        Obtain features and logits: \\
        $(\bm{z}_p^t, \bm{l}_p^t) \leftarrow f_{\theta_T^*}(\bm{p}^t)$, \\
        $(\bm{z}_e^t, \bm{l}_e^t) \leftarrow f_{\theta_S}(\bm{e}^t)$ \;
        
        Calculate $\mathcal{L}_{kd}$ and $\mathcal{L}_{align}$ by Eq. \eqref{loss_kd} and Eq. \eqref{loss_align} \;
        
        Update $\theta_S \leftarrow \nabla(\mathcal{L}_{kd} + \alpha \mathcal{L}_align)$ by Eq.\eqref{eq:loss_S} \;
    }
    
    Predict labels on the target sEMG feature $\{\bm{e}_j\}_{j=1}^{N_t}$ \;
    
    \KwRet Classification results $\{y_j\}_{j=1}^{N_t}$.
\end{algorithm}

\section{Experiment and Result}
\subsection{Datasets Description}
\begin{figure}
    \centering
    \includegraphics[width=1\linewidth]{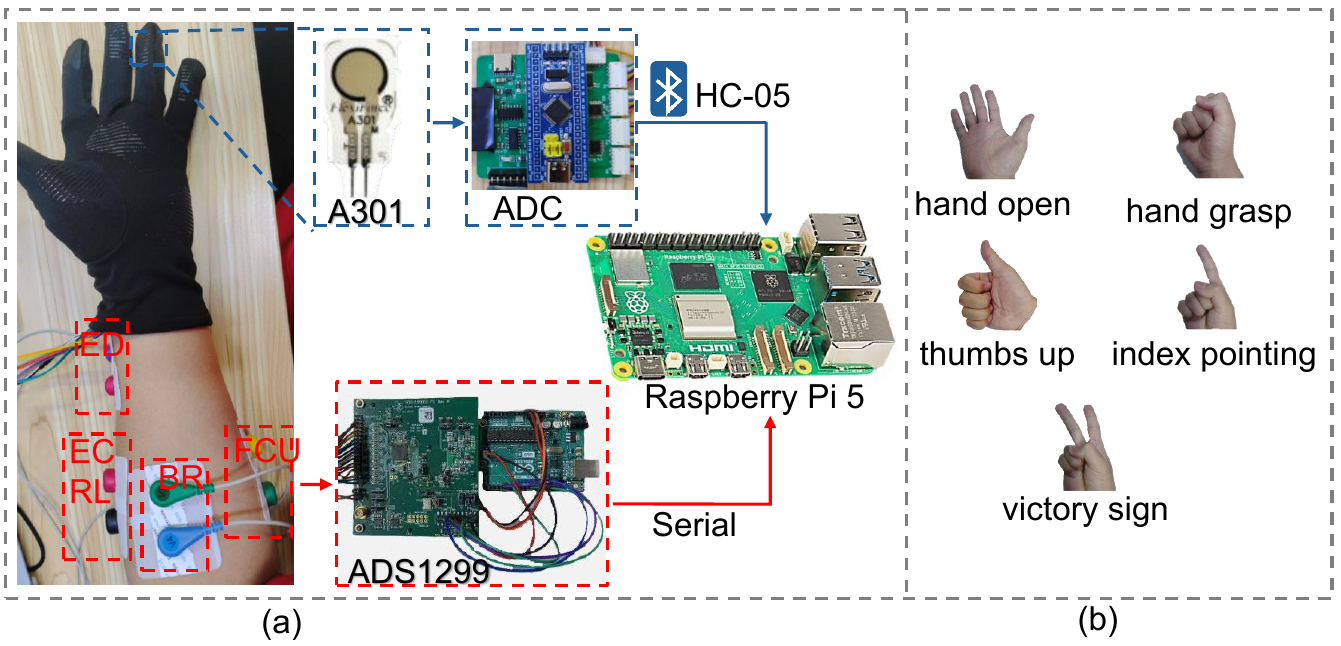}
    \caption{Data acquisition details. (a) The hardware architecture system; (b) Five selected hand gestures.}
    \label{fig:datacollect}
\end{figure}

To evaluate the effectiveness of the proposed PGUDA framework, we constructed a multimodal hand-gesture dataset containing synchronized sEMG and pressure signals.  
Data were collected from eleven healthy participants (21-28 years old, five males and six females, all right-handed) with no history of neuromuscular disorders. Each participant was asked to write informed consent prior to the experiment.
As illustrated in Fig. \ref{fig:datacollect}\hyperref[fig:datacollect]{(a)}, participants sat comfortably wearing a custom-made data glove and an sEMG acquisition device \cite{Zhong2025}. We deployed a four‑channel sEMG system with bipolar electrodes placed over the brachioradialis (BR), extensor carpi radialis longus (ECRL), flexor carpi ulnaris (FCU), and extensor digitorum (ED). Nine electrodes were arranged in a standard bipolar configuration, with a single reference electrode attached to the humeral epicondyle. sEMG signals were sampled at 500 Hz using a high‑precision 24‑bit analog‑to‑digital converter (ADS1299, Texas Instruments) and transmitted to a Raspberry Pi-5 via a serial interface for storage and processing.
Simultaneously, five force sensors (FlexiForce A301) were embedded in the fingertips of the custom glove. The pressure signals were amplified and digitized, then transmitted to the Raspberry Pi via an HC-05 Bluetooth module at 5 Hz.

The experiment included five hand gestures: hand open, hand grasp, thumbs up, index pointing, and victory sign, as shown in Fig. \ref{fig:datacollect}\hyperref[fig:datacollect]{(b)}. Each trial lasted 12 seconds, including 3s of preparation, 5s of gesture holding, and 4s of relaxation. Each gesture was repeated 10 times per session. Ten subjects participated in a single session (50 trials per subject), while one subject completed five separate sessions on different days (250 trials in total). To ensure high‑quality inputs, dedicated preprocessing pipelines were applied to each modality. For sEMG signals, a third‑order Butterworth band‑pass filter (3-120 Hz) removed motion artifacts and high‑frequency noise, and a 50 Hz notch filter suppressed power‑line interference. For pressure signals, given variations in sensor sensitivity, we applied min‑max normalization to scale each channel’s amplitude to the range $[0,1]$, ensuring numerical stability during model training.

\subsection{Implementation Details}
To thoroughly evaluate the proposed framework, we designed comprehensive unsupervised domain adaptation experiments covering both cross-subject and cross-session scenarios. For cross-subject evaluation, we adopted a pairwise transfer protocol among the 10 single‑session subjects: each subject served as the labeled source domain, while every other subject was treated as the unlabeled target domain, resulting in 90 transfer tasks. For cross-session evaluation, we utilized the data from the subject with five recording sessions. By selecting one session as the source and another as the target in a pairwise manner, we constructed 20 transfer tasks. With the adopted segmentation strategy, each session contained approximately 250 samples per gesture class.
The PGUDA framework was implemented in Python 3.11 using the PyTorch 2.8.0 library. All experiments were conducted on a platform equipped with an Intel Core i9-14900HX CPU and an NVIDIA GeForce RTX 5060 GPU. Network parameters were optimized using the AdamW algorithm, and the CosineAnnealingLR scheduler was employed to adjust the learning rate during training. Detailed hyperparameters settings are listed in Table \ref{tab:parameters}.

\begin{table*}[htbp]
\small
\centering
\caption{Classification Performance (Accuracy ± STD \% and $F_1$-score ± STD \%) \\ in Cross-Subject and Cross-Session evalution with Different Methods}
\label{tab:Comparison}
\setlength{\tabcolsep}{6pt} 
\begin{tabular}{ccccc}
\toprule
\multirow{2}{*}{Method} & \multicolumn{2}{c}{Cross-Subject}  & \multicolumn{2}{c}{Cross-Session} \\ \cmidrule{2-3} \cmidrule{4-5}
                          &  Acc ± Std & $F_1$ ± Std & Acc ± Std & $F_1$ ± Std                 \\ \midrule
KNN                       &  25.95 ± 8.84   & 22.44 ± 8.83    &  40.39 ± 7.55   & 37.60 ± 8.81  \\
MMD \cite{MMD2012}        &  34.16 ± 8.32   & 28.17 ± 9.68    &  47.86 ± 8.07   & 46.43 ± 8.59  \\
CORAL \cite{Sun2017}      &  34.53 ± 8.68   & 28.58 ± 9.68    &  48.34 ± 7.56   & 46.28 ± 8.98  \\
DANN  \cite{DANN2016}     &  36.94 ± 9.60   & 31.87 ± 10.69   &  49.08 ± 7.90   & 46.95 ± 10.06  \\
DSAN  \cite{DSAN2021}     &  36.05 ± 9.41   & 29.53 ± 9.91    &  47.64 ± 7.25   & 45.84 ± 7.79 \\
BNM   \cite{BNM2020}      &  36.45 ± 9.12   & 29.94 ± 10.28   &  47.75 ± 7.27   & 45.99 ± 7.42 \\
CDAN  \cite{Long2017ConditionalAD}   &  37.89 ± 9.65   & 32.95 ± 9.87   &  48.88 ± 7.41   & 47.42 ± 8.47\\
MCD   \cite{MCD2018}      &  39.03 ± 10.29  & 33.30 ± 10.83   &  50.75 ± 8.77   & 47.49 ± 10.96 \\
MDD   \cite{MDD2019}      &  41.14 ± 4.31   & 33.07 ± 6.10  &  45.64 ± 8.01   & 42.84 ± 9.67 \\

\textbf{PGUDA}    & \textbf{58.08 ± 17.84} & \textbf{54.21 ± 20.30}  &  \textbf{58.08 ± 10.73}   & \textbf{55.31 ± 12.27}  \\

\bottomrule
\end{tabular}
\end{table*}

\begin{table}[t]
\caption{Hyper-parameters Used During Training}
\label{tab:parameters}
\begin{tabular}{ccccccc}
\toprule
learning rate & $E_1$/$E_2$   & batch size & drop out & $T$   & $\alpha$ & $\mu$ \\ \midrule
0.003/0.00001  & 200/200 & 256        & 0.3      & 1.2 & 5  & 15 \\
\bottomrule
\end{tabular}
\end{table}

To validate the effectiveness of the PGUDA framework, we performed comparative evaluations against a comprehensive set of existing DA methods, which are summarized as follows:
\begin{itemize}
    \item MMD \cite{MMD2012}: A distance-based method that minimizes the distribution discrepancy between the source and target domains in a Reproducing Kernel Hilbert Space.
    \item CORAL \cite{Sun2017}: A statistical method that aligns the second-order statistics (covariance matrices) of the source and target feature distributions.
    \item Domain-adversarial neural network (DANN) \cite{DANN2016}: A pioneering adversarial method that employs a gradient reversal layer to confuse a domain discriminator, thereby learning domain-invariant features.
    \item Deep subdomain adaptation network (DSAN) \cite{DSAN2021}: An extension of MMD that aligns relevant subdomain distributions based on local MMD.
    \item Batch nuclear-norm maximization (BNM) \cite{BNM2020}: An unsupervised method that maximizes the nuclear norm of the batch output matrix to enhance the discriminability and diversity of predictions on the target domain.
    \item Conditional domain adversarial network (CDAN) \cite{Long2017ConditionalAD}: A conditional adversarial method that conditions the adversarial adaptation on discriminative information to capture multimodal structures.
    \item Maximum classifier discrepancy (MCD) \cite{MCD2018}: A method that utilizes a min-max adversarial game between a feature generator and two classifiers to align distributions by detecting target samples near the decision boundaries.
    \item Margin disparity discrepancy (MDD) \cite{MDD2019}: A theory-driven method that provides rigorous generalization bounds by minimizing margin disparity.
\end{itemize}

To establish a performance benchmark, we employ the K-Nearest Neighbors (KNN) classifier. The evaluation is conducted using accuracy and $F_1$-score. All experimental results are reported in the form of $\text{average accuracy} \pm \text{standard deviation}$.

\subsection{Experimental Results}
\subsubsection{Performance on Cross-Subject Task}
The cross‑subject experimental results are summarized in the left columns of Table \ref{tab:Comparison}. This setting is particularly challenging owing to inherent anatomical variations across individuals, such as muscle geometry and skin impedance.
Consequently, the baseline KNN model performs poorly, achieving average accuracies of only 25.95\%, which reflects the substantial domain discrepancy. Although conventional domain adaptation methods (e.g., CORAL and DANN) yield modest improvements, they are still limited by the complex distribution shifts characteristic of sEMG signals.
By contrast, the proposed PGUDA framework demonstrates strong generalization. It surpasses the second‑best method by 16.94\% in accuracy and 20.91\% in $F_1$-score, indicating its ability to extract subject‑invariant representations and effectively mitigate the impact of inter‑subject physiological differences.

\subsubsection{Performance on Cross-Session Task}
We further evaluate the longitudinal robustness of the PGUDA using the cross-session protocol, as presented in the right columns of Table \ref{tab:Comparison}. 
Unlike cross‑subject scenarios, cross‑session adaptation faces challenges from temporal non‑stationarity and electrode displacement over time.
In this setting, both the baseline KNN and DA methods like DANN and CDAN exhibit high variance, largely due to the instability of min-max optimization on time-varying distributions. However, PGUDA demonstrates superior robustness, outperforming the best existing methods by 7.33\% in accuracy and 7.82\% in $F_1$-score.
This result confirms that the physical constraints provided by the pressure modality are inherently time-invariant. By transferring this physical stability to the sEMG network, our method effectively bridges the temporal shifts, verifying its suitability for long-term practical deployment.

\begin{figure*}[t] % [t] 表示强制置顶，跨栏图通常只能放在顶部
    \centering

    \includegraphics[width=1\linewidth]{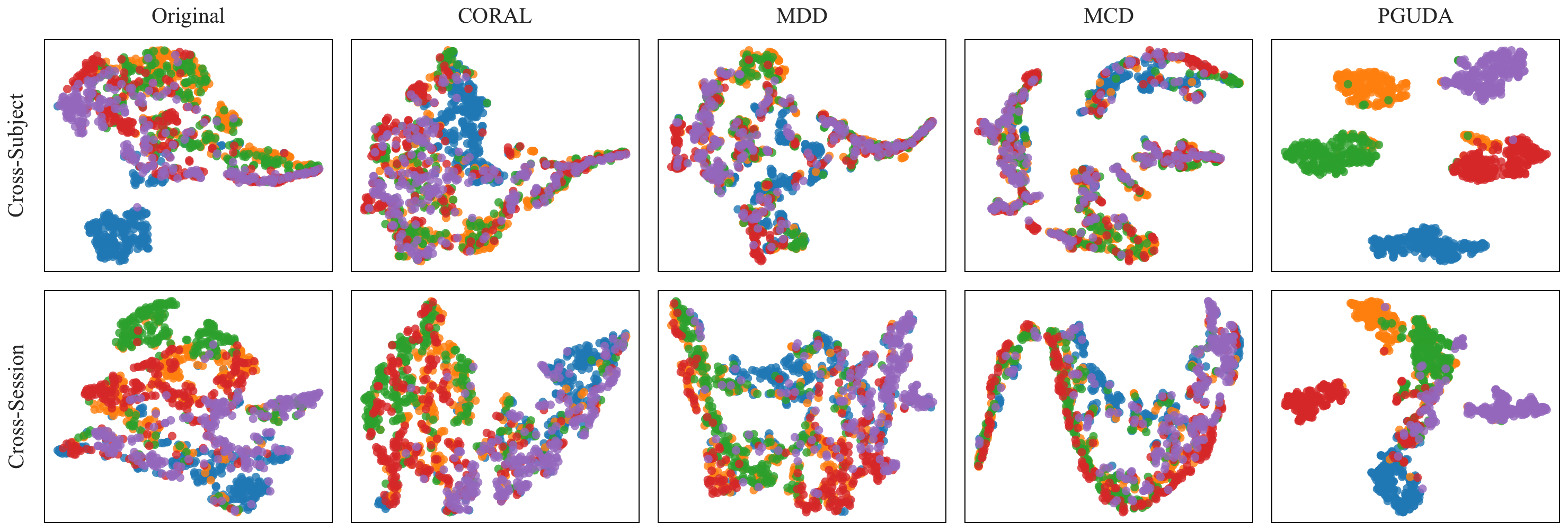}
    
    \caption{t-SNE visualization of feature distributions with different methods. The top row is the task of Subject 2 $\rightarrow$ Subject 3, and the bottom row is the task of Session 4 $\rightarrow$ Session 2. Blue, yellow, green, red, and purple samples denote the hand gestures of hand open, hand grasp, thumbs up, index pointing, and victory sign, respectively.}
    
    \label{fig:tsne_comparison}
\end{figure*}

\subsection{Featuer Visualization}
Fig. \ref{fig:tsne_comparison} presents the t-SNE visualization of the feature distributions for CORAL, MCD, MDD, and the proposed PGUDA in both cross-subject and cross-session evaluations. In the original feature space, sEMG features exhibit a chaotic distribution with substantial overlap among different gesture classes. This lack of separability confirms the severity of the domain shift and explains the poor baseline performance. Conventional DA methods (\textit{e.g.} CORAL, MDD, and MCD) attempt to align the domain distributions but struggle with complex classes. For instance, the "victory sign" gesture remains difficult to classify due to persistent overlap. While those methods reduce the confusion to some extent, the resulting decision boundaries are often unclear. In contrast, our PGUDA framework produces notably more discriminative features. Although PGUDA does not enforce strict geometric overlap between source and target domains, it successfully projects the target sEMG data into well‑separated, structured clusters. The visualized features exhibit high intra‑class compactness and clear inter‑class separation. This indicates that by leveraging the physical stability of pressure as semantic guidance, PGUDA enables the model to learn robust decision boundaries that are insensitive to subject‑specific variations, rather than merely aligning marginal distributions.

\section{Discussion}

\begin{figure*}[t] % [t] 表示强制置顶，跨栏图通常只能放在顶部

    \includegraphics[width=1\linewidth]{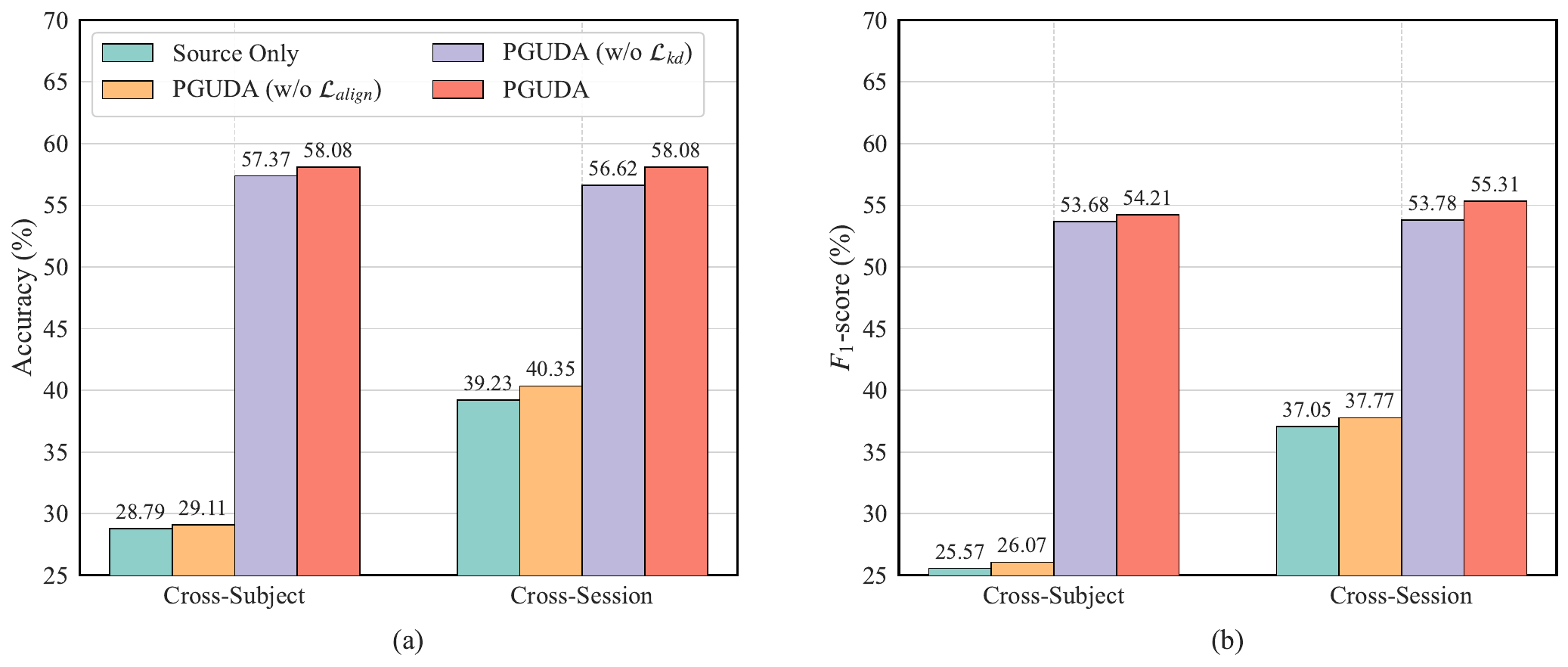}
    
    \caption{Ablation study results of the PGUDA. Subplots (a) and (b) illustrate the classification Accuracy and $F_1$-score across cross-subject and cross-session tasks. The "Source Only" means the baseline model trained solely labeled source data without any adaptation; The terms PGUDA (w/o $\mathcal{L}_{align}$) and PGUDA (w/o $\mathcal{L}_{kd}$) denote variants where the feature alignment loss and knowledge distillation loss are removed. The full PGUDA model consistently achieves the highest performance, validating the contribution of each component.}
    
    \label{fig:ablation}
\end{figure*}

\subsection{Ablation Experiment}
To systematically investigate the contribution of each component in the PGUDA framework, we conduct ablation studies under both cross-subject and cross-session protocols. As shown in Fig. \ref{fig:ablation}, we compare the full PGUDA model with three variants: the "Source Only" baseline, PGUDA trained without teacher guidance (w/o $\mathcal{L}_{kd}$), and PGUDA trained without feature alignment (w/o $\mathcal{L}_{align}$).
The results provide clear insights into the mechanisms underlying the student network's learning behavior.

A key observation is that PGUDA (w/o $\mathcal{L}_{kd}$) yields marginal improvements over the "Source Only" baseline. Specifically, under the cross-subject protocol, the classification accuracy increases slightly from 28.79\% to 29.11\%, while in the cross-session task, the accuracy remains nearly unchanged at approximately 40\%.
This indicates that, given the high stochasticity and complexity of sEMG signals, merely matching the statistical distributions of the pressure and sEMG data is insufficient. Without reliable semantic anchors, feature alignment may incorrectly map samples from different classes, failing to establish correct decision boundaries.

In contrast, PGUDA (w/o $\mathcal{L}_{align}$) achieves a substantial performance gain, with accuracy rising to 57.37\% in cross-subject and 56.62\% in cross-session tasks. This highlights the knowledge distillation (KD) module as the primary driver of improvement.
Our finding confirm the core hypothesis: the pressure modality is physically robust and domain-invariant. By mimicking the soft logits from the pressure-based teacher, the student network effectively inherits latent knowledge about inter-class relationships. This semantic guidance acts as a strong regularizer, enabling the student to classify target sEMG patterns correctly even without explicit feature distribution constraints.
Finally, the PGUDA framework achieves the best performance across all variants. Integrating $\mathcal{L}_{align}$ with $\mathcal{L}_{kd}$ provides additional improvements over KD alone, indicating that the two components are complementary.
KD supplies discriminative guidance to correct decision boundaries, while feature alignment further compacts the intra-class representations and reduces global domain shift. Together, they ensure both semantic correctness and distributional consistency.

\begin{figure}[] % [t] 表示强制置顶，跨栏图通常只能放在顶部

    \includegraphics[width=1\linewidth]{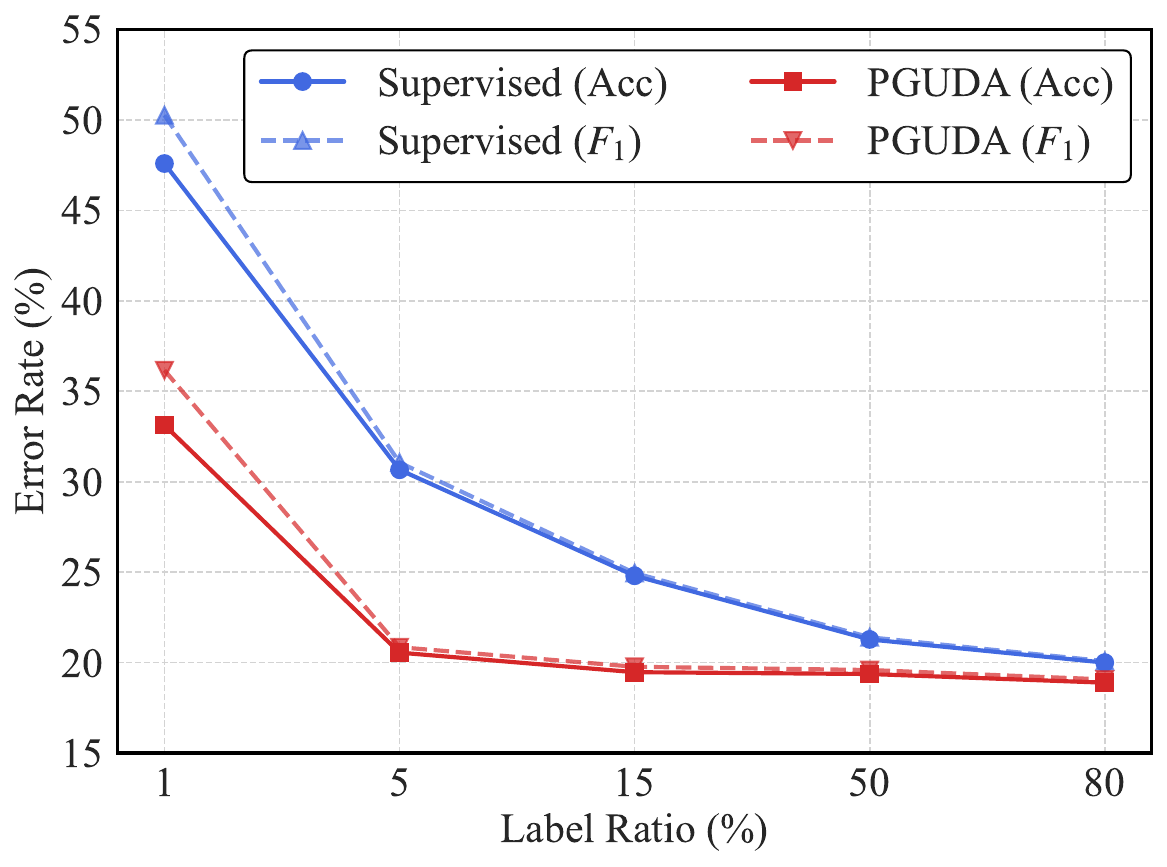}
    
    \caption{Error rate comparison (Accuracy and $F_1$-score) under different labeled data ratios. The proposed PGUDA (red) consistently achieves lower error rates than the Supervised baseline (blue) across all settings, demonstrating high label efficiency and robust generalization.}
    
    \label{fig:error_rate}
\end{figure}

\subsection{Label Efficiency Analysis}
In practical sEMG applications, acquiring large-scale annotations is labor-intensive and time-consuming. To evaluate the label efficiency of our PGUDA framework, we conduct a semi-supervised learning experiments under label scarcity using intra-subject 5-fold cross-validation.
We randomly sample subsets of the dataset (1\%, 5\%, 15\%, 50\% and 80\%) as the labeled set, treating the remaining data as unlabeled. The experiment compares two paradigms: (1) a supervised baseline trained directly on sEMG features with sparse labels, and (2) PGUDA, in which the sparse labels are used only to train the teacher model, while the sEMG student network is trained via cross-modal distillation from synchronized pressure-sEMG pairs without explicit sEMG annotations.

The results are summarized in Fig. \ref{fig:error_rate}.
As can be seen, the error rate of both methods degrades sharply as the label ratio increases. Our proposed PGUDA outperforms the supervised baseline across all label ratios. Even when both methods use the full set of labels (80\%), PGUDA still achieves a lower error rate than the supervised baseline.
Notably, PGUDA with only 5\% labeled data attains comparable performance to the supervised baseline trained with 80\% labels.
Specifically, PGUDA reaches an error rate of 20.55\% with 5\% labels, while the baseline achieves 20.00\% with 80\% labels, a marginal difference of only 0.55\%. This demonstrates the high label efficiency of PGUDA.
Furthermore, under an extremely low label ratio (1\%), PGUDA still significantly reduces the error rate (from the baseline’s 47.59\% down to 33.13\%), illustrating its strong few-shot learning capability. As the label ratio increases, PGUDA consistently maintains a superior error rate, further validating the robustness and generalization ability of the proposed method.

% In contrast, PGUDA demonstrates exceptional resilience to label scarcity:
% \begin{itemize}
%     \item At 1\% labels: PGUDA achieves an accuracy of 66.87\%, outperforming the supervised baseline by a significant margin of 14.46\%. This indicates that the cross-modal knowledge distilled from the pressure teacher effectively compensates for the lack of explicit sEMG supervision.
%     \item At 5\% labels: Remarkably, PGUDA attains an accuracy of 79.45\%. This performance is comparable to the fully supervised upper bound (80.00\% accuracy using 80\% labels). This implies a profound practical advantage: users only need to annotate 5\% of the pressure data to train a robust sEMG model that performs nearly equivalently to a fully supervised system trained on massive datasets.
%     \item At 15\% and 80\% labels: With 15\% labels, PGUDA reaches 80.53\%, already surpassing the 80\%-data supervised upper bound. When the label ratio is increased to 80\%, performance marginally improves to a peak of 81.11\%. While this gain is incremental compared to the 15\% setting, it confirms two critical findings: first, PGUDA consistently outperforms the supervised baseline regardless of data volume; second, and more importantly, our framework is highly data-efficient, achieving near-optimal performance without requiring extensive annotations.

% \end{itemize}

% These findings highlight PGUDA as a highly data-efficient learner, capable of delivering robust performance with minimal annotation effort.

\subsection{Parameter Sensitivity Analysis}
\begin{figure}
    \centering
    \includegraphics[width=1\linewidth]{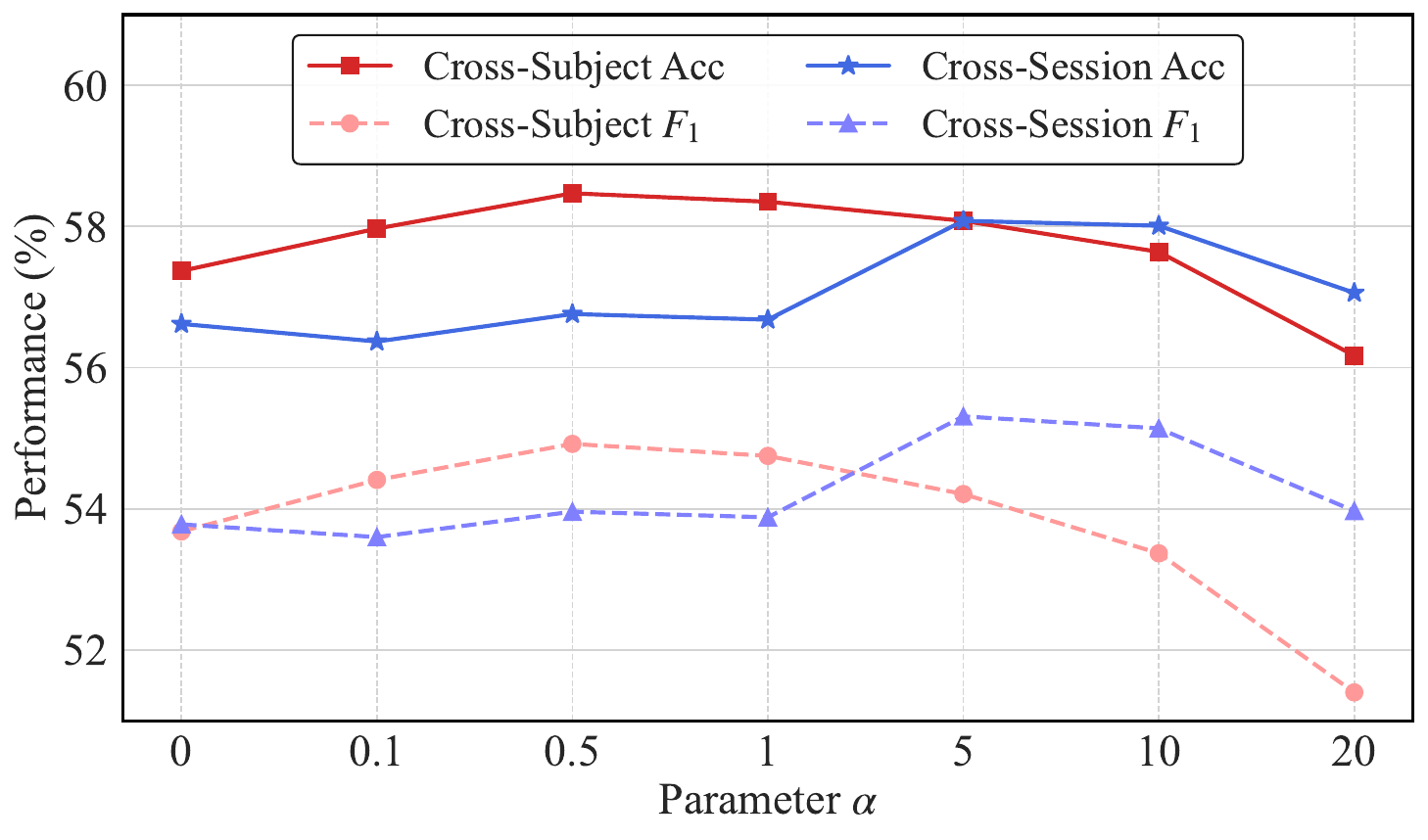}
    \caption{Impact of the trade-off hyperparameter $\alpha$ on model performance. The dual-axis plot displays Accuracy  and $F_1$-score  for both cross-subject and cross-session protocols.}
    \label{fig:parameter_sensitivity}
\end{figure}

The hyper-parameter $\alpha$ in Eq. \eqref{eq:loss_S} controls the trade-off between the knowledge distillation and the feature alignment loss. A larger $\alpha$ emphasizes stricter cross-domain distribution matching, while a smaller $\alpha$ prioritizes preserving of discriminative features learned from the source domain.
To evaluate its impact, we varied $\alpha$ across $\{0, 0.1, 0.5, 1, 5, 10, 20\}$ in both cross-subject and cross-session tasks. The results are presented in Fig. \ref{fig:parameter_sensitivity}.
In cross-subject tasks, the model performance remains robust and consistently superior to the baseline ($\alpha=0$) across a wide range of $\alpha$ values $[0.1, 10]$, peaking within $\alpha \in [0.5, 5]$. The accuracy peaks around $\alpha \in [0.5, 5]$. However, when $\alpha$ increases to 20, accuracy and $F_1$-score both declines, indicating that excessive alignment weight may induce negative transfer.  Overly aggressive distribution alignment can distort the intrinsic structure of sEMG features, thereby degrading the separability of gesture categories. 
Cross-session tasks exhibits a different sensitivity pattern. Compared with the no-alignment case, using low $\alpha$ values ($0.1$ and $1$) leads to only marginal performance gains. Significant performance gains emerge only when $\alpha$ reaches $[5, 10]$. 
This suggests that temporal signal drifts are more persistent or subtle than cross-subject variations, thus requiring stronger regularization to align the time-varying target distribution with the source domain. Based on these observations, we set $\alpha=5$ as a balanced and robust setting suitable for both scenarios.

\subsection{Why Cross-modal Knowledge Distillation Work?}
Experimental results confirm that the cross-modal knowledge distillation (CMKD) module is decisive for the success of the PGUDA framework, significantly outperforming pure feature alignment methods. The effectiveness of CMKD stems from three key factors rooted in physiological invariance, latent knowledge transfer and semantic guidance for domain adaptation.
The primary challenge in sEMG-based gesture recognition is the high variability of bio-electrical signals caused by domain shift. In contrast, pressure signals capture the mechanical outcome of hand gestures. Regardless of arm morphology or skin condition, the mechanical contact pattern of a "hand grasp" gesture remains physically consistent across domains. By using the pressure-based model as the teacher, we introduce a domain-invariant physical constraint into the learning process. The distillation forces the sEMG student network to discard domain-specific noise and concentrate on latent features that correlate causally with stable mechanical outputs.

In contrast to the categorical hard labels used in supervised learning, the soft logits produced from the teacher encode rich latent knowledge about inter-class relationships. For example, when recognizing an “index pointing” gesture, the teacher may also assign a non‑zero probability to the “victory sign” due to similarities in finger extension. This soft probability distribution provides a granular description of the decision boundary manifold. By mimicking these soft targets, the student learns not only the correct class but also the relative affinities among classes. This acts as a regularizer that mitigates overfitting to the source domain’s specific distribution.
Our ablation study reveals that pure feature alignment fails to deliver meaningful improvement. When the domain gap is too large, as in cross-subject tasks, indiscriminate minimization of distribution distance may align samples from different classes and degrade discriminability. The CMKD module provides essential semantic guidance: the teacher generates reliable pseudo-supervision for unlabeled target data, effectively anchoring target features to their correct semantic clusters before the alignment loss further compacts them. In short, CMKD ensures alignment correctness, while feature alignment ensures representational compactness.

\section{Conclusion}
This study addresses the challenges of domain discrepancy and label scarcity in sEMG-based gesture recognition by proposing the PGUDA framework. By leveraging the physical stability of pressure signals to guide sEMG feature learning, PGUDA effectively extracts domain-invariant representations across subjects and sessions. Comprehensive evaluations under cross-subject and cross-session protocols demonstrate that our proposed method significantly outperforms existing domain adaptation approaches, achieving an overall accuracy of 58.08\%. Ablation studies verify that the cross-modal knowledge distillation mechanism serves as the primary driver of performance improvement, effectively preventing negative transfer by transferring physically consistent semantic knowledge. Furthermore, the framework exhibits exceptional label efficiency, achieving performance comparable to supervised benchmarks while utilizing only 5\% of labeled pressure data, thereby substantially reducing the calibration burden for new users. Future work will focus on implementing online incremental learning for real-time edge adaptation and extending the framework to continuous force regression tasks.

\section*{Acknowledgments}
The authors would like to thank all the subjects who participated in this study. We hereby declare that written informed consent was obtained from all participants prior to the experiments, and all personal data have been anonymized for privacy protection.

%\begin{refcontext}[sorting = none]
%\printbibliography
%\end{refcontext}%

\bibliographystyle{IEEEtran}
%\bibliography{Bibliography}

\vfill

\end{sloppypar}
\end{document}